# Extreme bright coherent synchrotron radiation produced in a low emittance electron storage ring by the angular dispersion induced microbunching scheme

Changliang Li[1,2,3], Chao Feng[1,2], Bocheng Jiang[1,2,*]

[1]Shanghai Advanced Research Institute, Chinese Academy of Sciences, Shanghai 201210, China
[2]Shanghai Institute of Applied Physics, Chinese Academy of Sciences, Shanghai 201800, China
[3]University of the Chinese Academy of Sciences, Beijing 100049, China

## ABSTRACT

Generation of extreme bright coherent synchrotron radiation in a short wavelength range is of remarkable interest in the synchrotron light source community. In this paper, a novel technique is adopted to produce the coherent radiation which uses an angular dispersion to enhance the micro-bunch by a tiny amplitude of energy modulation. This scheme can be inserted in a long straight section of a storage ring. A lattice design of an extreme-low emittance storage ring with 3.5 GeV energy is presented in this paper, which employs a higher order achromat (HOA) concept. We show the design results for the multi-bend achromat (MBA) lattice with an emittance of 20 pm-rad in about 900-meter circumference. Numerical simulations demonstrated that this angular dispersion induced micro-bunch scheme can be used to generate coherent radiation near soft x-ray region, the repetition rate can reach 10 kHz and the spectral brightness can reach $3 \times 10^{24}$ phs/s/mm$^2$/mrad$^2$/0.1%BW.

## I. INTRODUCTION

The development of advanced particle accelerator technology, especially the emergence of synchrotron radiation light source and free electron laser (FEL), has brought unprecedented revolutionary research tools to the fields of physics, chemistry, life science, material science and greatly promoted the development of related science [1]. The storage ring based synchrotron radiation source, which was born in the 1960s, has undergone three generations of development and evolution, and is currently moving towards the fourth generation aiming at diffraction-limited storage rings with higher brightness and better transverse coherence. Storage ring based synchrotron radiation sources have now become a major scientific platform supporting multidisciplinary development of basic and applied research. The synchrotron light source has many advantages such as wide spectral coverage, high average brightness, high stability, and simultaneous support for multi-users [2].

Linear accelerator-based FEL is an advanced light source. FEL not only has ultra-high peak brightness, ultra-short pulse structure and longitudinal coherence, but also has the ability to continuously adjust the wavelength of the light [3-5]. The emergence of such EUV and X-ray FEL with short time, high energy and spatial resolution provides unprecedented tools for detecting the ultra-fast evolution of microstructures. One issue of FEL is pursuing to increase the repetition rate of the radiation. High-frequency electron gun and superconducting linear accelerator technology are used for high repetition FEL production, while the cost and technical difficulty is great. In addition, FEL devices generally can only supply the light to a limited number of experimental stations, which also limits its application scope to a certain extent.

*jiangbocheng@zjlab.org.cn

As bunch length in the storage ring is relatively longer than in the linac, which limits the temporal coherent production. Several attempts have been tried in the storage ring to produce temporal coherent [6,7]. The main idea is to combine storage ring with FEL. There are two main modes of operation: one is a low-gain FEL based on an optical resonator, the other is coherent harmonic generation technologies (CHG-FEL). Resonant cavity FEL is limited by reflective materials, and the output wavelength is difficult to reach below 200 nm. CHG-FEL requires the use of a conventional laser as a seed and its harmonic conversion to produce short-wavelength radiation. The storage ring-based CHG-FEL seems to provide users with an ideal ultra-short pulse full-coherent light source. However, the number of harmonic conversions of CHG-FEL is limited by the energy spread of the electron beam, and the number of harmonics that can be reached is generally very limited. The typical electron beam energy spread in the storage ring is $10^{-3}$. To generate a certain amount of harmonic bunching, on the one hand, a super-strong seed laser (with a peak power in the range of GW to 100 GW) needs to introduce enough energy modulation, on the other hand, limits the depth of energy modulation to ensure that the quality of the electron beam is not degraded. It is precisely because of many practical limitations that in the past two decades, CHG-FEL based on storage ring has not achieved great development. Meanwhile, femto slicing method has been proposed to achieve femtosecond radiation pulses in storage ring-based light sources [8,9]. It employs the resonant interaction of an electron bunch with a femtosecond laser beam in a wiggler to energy-modulate a short section of the bunch. The induced energy modulation is then converted to a transverse displacement using a vertical dispersion bump downstream of the wiggler. Thus, the radiation from the femtosecond pulse can be separated from the main bunch radiation. The repetition rate of the femto slicing can reach as high as MHz level.

Storage ring synchrotron radiation hold the property of high repetition rate and relatively low peak power. FEL has low repetition rates yet extremely high peak power. Among the many applications for high-power coherent radiation sources, some do not demand high peak power. Instead, they focus on having a high average-power and high repetition rate. To approach high average power radiation, steady state micro-bunch (SSMB) had been proposed since 2010 [10]. SSMB is based on an electron storage ring which is much more mature comparing to the energy recovery linac (ERL), the latter one is also a candidate for high average power radiation provider yet under developing [11]. When micro-bunch length closes to the radiation wavelength, coherent radiation will be produced and radiation power will be orders of magnitude higher, together with high repetition rate, the high average power radiation will be produced. Beyond the scientific applications, it will have some important industry applications such as EUV lithography.

At present, two kinds of SSMB scenarios are under profound study, namely strong focusing SSMB and reversible SSMB [12-15]. Either of them encounters several challenges. On the way of pursuing reversible SSMB, we found that to produce turn-by-turn and bunch-by-bunch coherent radiation not only requires a novel lattice design to reduce high order terms to an extraordinary low level, but also requires very stable lasers for energy modulation and re-modulation to cancel each other makes beam transparent to the rest part of the storage ring. Reducing the repetition rate of reversible SSMB will highly increase the feasibility. In this paper, we will show that it is possible to produce 10 kHz coherent radiation without re-modulation. Without re-modulation, the challenges will be greatly reduced. The energy modulated beam needs to damp to the ground state by radiation damping which limits the repetition rate. However with 10 kHz repetition rate, the average brightness is still 2 orders of magnitude higher than the fourth generation synchrotron light source. Such high bright coherent synchrotron radiation will benefit synchrotron users in many aspects.

The paper is organized in the following structures. In Sec. II, the angular dispersion induced micro-bunching scheme is described and related numerical simulations are presented. In Sec. III, we present the lattice design, linear optics and non-linear optics. In Sec. IV, beam distortion, repetition rate and brightness are analyzed. Finally, the discussions and conclusions are given in Sec. V.

## II. ANGULAR DISPERSION INDUCED MICRO-BUNCHING SCHEME

The angular dispersion induced micro-bunching (ADM) [16-18] is a suitable method to produce high harmonic by a very weak modulate when an electron beam gets very small vertical emittance which is naturally preserved in an electron storage ring. We recall this scheme in Ref. [15], the layout of it as shown in Fig. 1. A magnetic dipole (B) is added upstream of the modulator (M), following a dogleg that consists of two dipoles of opposite polarity as the dispersion section (D). The first dipole is used to introduce an angular dispersion into the electron beam. Then seed laser pulse at optical wavelength is employed to interact with the electron beam in the modulator to introduce a small energy modulation. After that, the energy modulation is converted into density modulation by the dogleg. The dispersive properties of the first dipole and the dogleg allow, if the parameters are chosen properly, the full compensation of the initial beam energy spread to produce very sharp micro-bunches at the EUV wavelength. This kind of electron beam would help to initiate intense coherent radiation at EUV wavelength in the following undulator. This scheme can be inserted in a long straight section of the ring. The transverse dispersion generated by the dogleg can be fully compensated by another reversed dogleg after the radiator (R). This technique can make full use of the low emittance of the electron beams of a storage ring while effectively mitigating the detrimental effect from the large energy spread in the meantime.

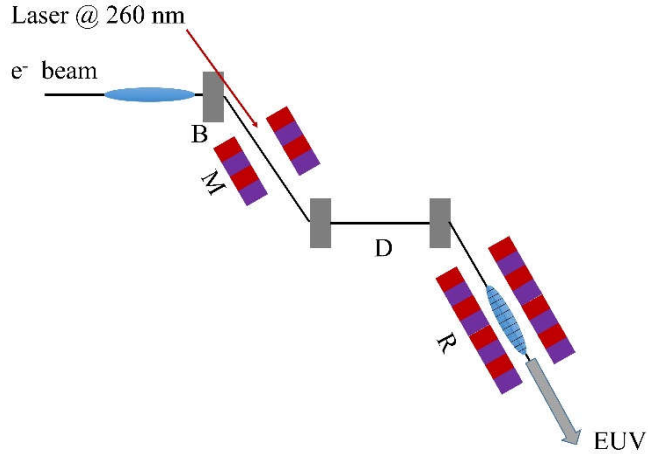

Fig. 1. Layout of the angular dispersion induced micro-bunching scheme.

Detailed matrix derivation and bunching factor analysis can be referred to Ref. [16]. Here we give out numerical tracking results. The high harmonic bunching factor is determined by the product of the dispersion generated in the dogleg and the initial beam vertical divergence. In order to enhance the bunching factor for a given energy modulation amplitude, one can increase the beta function at the entrance of the first dipole to reduce beam vertical divergence or increase the strength of the first dipole to reduce the required dispersion of the dogleg.

Three-dimensional (3D) numerical simulations are necessary to show the possible performance of the scheme. The electron beam is tracked through the beam line using the code ELEGANT with second-order transport effects taken into account [19]. The detailed ring lattice design will be shown in Sec. III, here we only list the beam parameters used in the 3D simulations in Table I. Where, H and V denote the horizontal and vertical directions, respectively.

TABLE I.  Main Parameters used in 3D simulations

| Parameters | Values |
|---|---|
| Energy [GeV] | 3.5 |
| Energy spread [MeV] | 3.6 |
| Emittance (H, V) [pm rad] | (20, 20) |
| Laser wavelength [nm] | 260 |
| Laser power [MW] | 200 |
| Energy modulation amplitude [MeV] | 1.2 |
| Beta function at B (H, V) [m] | (10, 28) |
| Alpha function at B (H, V) [m] | (0.5, 0) |
| Length of dipole magnets [m] | 0.3 |
| Bending Angle of the first B [mrad] | 11.8 |
| Bending Angle of the dipole in D [mrad] | 14.8 |
| Distance between two dipoles in D [m] | 0.35 |
| Total Length [m] | 4.4 |

The first dipole creates a correlation between the angular divergence and energy deviation of the electron beam. As the initial angular divergence of the electron beam is very small, the angular divergence represents the electron energy deviation after passing through the first dipole. The initial longitudinal phase space is shown in Fig. 2(a). After passing through the first dipole with a length of 0.3 m and a bending angle of about 11.8 mrad, the electron beam is sent into a short modulator to interact with seed laser at wavelength 260 nm, and then the phase space becomes Fig. 2(b). This energy modulation is then transformed into an associated density modulation by a dogleg, shown in Fig. 2(c). The dipole magnets in the dispersion section has a length of 0.3 m and a bending angle of about 14.8 mrad. The distance between two dipoles in the dispersion section is 0.35 m.

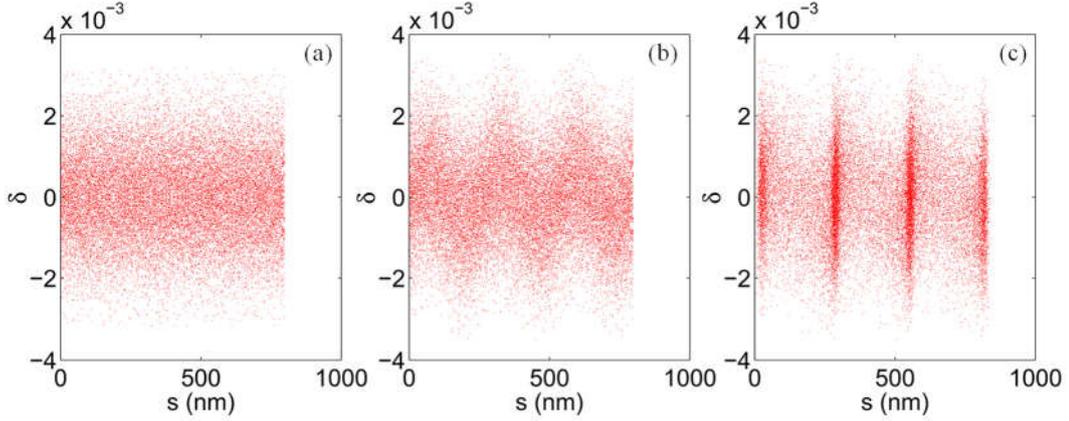

Fig. 2. The longitudinal phase-space evolution: (a) the initial phase space; (b) the phase space at the exit of the modulator; and (c) the phase space at the entrance of the radiator.

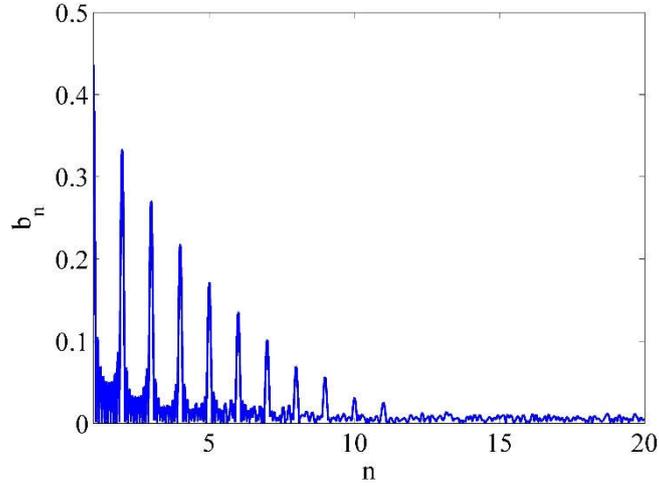

Fig. 3. The bunching factors of the ADM. The results are from 3D simulations for this scheme with an energy modulation amplitude of about 0.34 times of the initial beam energy.

With a very small energy modulation of about 0.34 times of the initial beam energy spread, one can find that the bunching factor is about 0.14 at the 6$^{th}$ harmonic (as shown in Fig. 3) which is at 44.3 nm coherent radiation. The bunching factor is sufficiently large to effectively suppress the electron beam shot noise and to generate coherent radiation in the following radiator.

## III. STORAGE RING DESIGN

The storage ring light sources currently running in the world are mostly third generation light sources, ADM can be adopted in the third generation light sources, yet for advanced radiation production, ADM based on a fourth generation light source is more attractive. ADM adopted in the fourth generation light source will increase the average brightness even 2 orders of magnitude. Furthermore, study ADM in the fourth generation light source will benefit the future investigation of reversible SSMB because an accurate reverse modulation needs an ultra-low emittance to reduce the high order effects. ADM requires an ultra-low vertical emittance to produce high harmonics, in the third generation light source transverse coupling can be controlled to a very small extent [20], but for long-term operation the change of transverse coupling degree of 0.1% has a great impact on the vertical emittance, which will affect the bunching factor. While for the fourth generation light source's the horizontal emittance is already extremely low, the vertical emittance is not sensitive to the transverse coupling. Based on these reasons, a lattice for the fourth generation of light source is used for ADM in this paper.

The fourth generation of light sources has been emerging based on the multi-bend achromat (MBA) lattice types [21,22]. The MBA lattice types exploits the inverse cubic dependence of emittance on the number of bending magnets. By choosing a very small bending angle per dipole, the emittance can be dramatically reduced. The synchrotron radiation source MAX IV in Lund, Sweden, is the first light source that was successfully commissioned with this new lattice type [23,24]. Hybrid multiband achromat (HMBA) is another effective way to decrease emittance. It was employed for the first time in the upgrade plan of ESRF [25]. In the Hybrid lattice cell, the sextupoles are located in the two dispersion bumps at both sides of the cell, with sextupoles of each family separated by a –I transformation to cancel part of their nonlinear effects. Many other sources worldwide follow this idea, such as APS-U [26], HEPS [27], and HALS [28,29].

As an alternative, a scheme known as Higher Order Achromat (HOA) develops a MBA lattice where chromaticity correcting sextupole magnets are distributed in each unit cell with strict phase advances over the cell such as to cancel basic geometric and chromatic resonance driving terms [30]. One important addition is the use of longitudinal-gradient bends (LGB) to suppress the dispersion at the LGB center, where the field is strongest [31,32]. Furthermore, the focusing quadrupoles can be offset radially in order to provide aggressive reverse bending (RB) [33] which allows tuning the dispersion independently from the beta functions and thereby optimizes the unit cell for ultra-low emittance and appropriate phase advance. SLS-2 is a typical example of using this scheme [34]. Since one of the most important parameters for the users of synchrotron radiation is brightness which is determined by the emittance in the storage ring, our aim in this section is to investigate and present the design results for a lattice which has extremely low emittance in 3.5 GeV which that can provide high brightness photons near soft X-ray diffraction limited. Our designed lattice is based on HOA concept.

Fig. 4 shows the optical functions for the unit cell for our design. The cell has a net bending angle of (10/6)°, which is made up from the 1.951734° deflection of the center LGB and the 2× (-0.142534°) deflection of two RBs. In Fig. 4 the two slices outside the LGB have a very small deflection angle and with defocused quadrupole field. In order to achieve a higher-order achromat across the seven cells, the phase advances for the unit cell are then (3/7, 1/7)×2π [35,]. The intensity of RB greatly affects the horizontal emittance and momentum compression factor $\alpha_c$. Fig. 5 shows the effect of RB at different intensities on horizontal emittance and momentum compression factor. The color bar represents the ratio of RB's deflection angle to the total net deflection angle. Our design goal is the emittance about 20pm-rad. Based on this goal, we choose the deflection angle of -0.142534°, and get $\alpha_c = -2.4e^{-5}$.

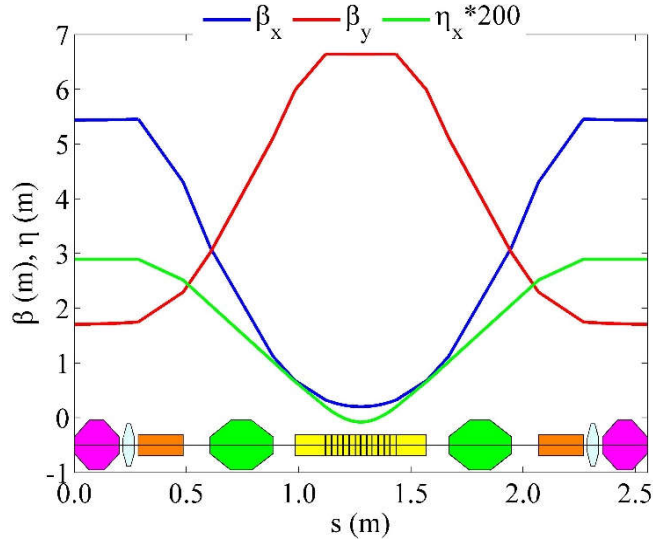

Fig. 4. Optical functions for the unit cell. LGB is in yellow rectangle, RB in red rectangle and sextupoles in green and magenta hexagon, octupole in white. The unit cell starts and ends at the half of sextupole magnets.

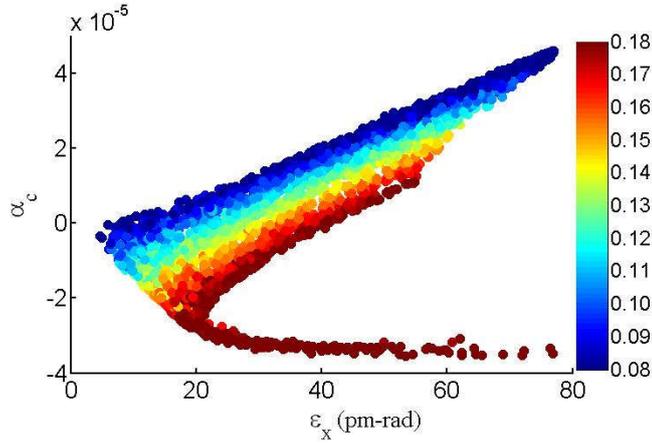

Fig. 5. Momentum compaction factor vs horizontal emittance for RBs with different strengths

The storage ring lattice consists of 36 seven-bend achromats separated by 4.7 m straight sections for insertion devices (ID). Each of the achromats consists of five unit cells and two matching cells. The unit cells have a (10/6)° bending magnet, while the matching cells at the ends of the achromat have a (10/12)° bending magnet. The optical functions of one cell are shown in Fig. 6. The maximum strength of dipole and quadrupole magnet is 1.6 T and 81.6 T/m, respectively. The horizontal and vertical tune of the 7BA one cell are set to $\mu_x = 19/6$ and $\mu_y = 7/6$ respectively for the nonlinear cancellation [36].

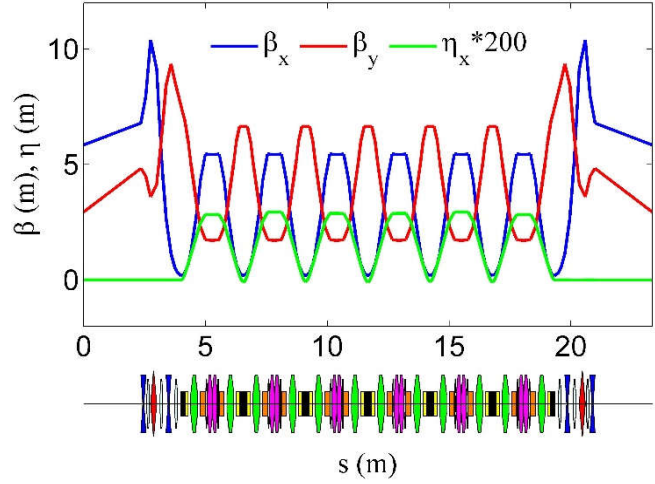

Fig. 6. Optical functions for the one cell. The yellow, red, blue, green, magenta and white indicate dipole, focus quadrupole, defocus quadrupole magnets, focus sextupole, defocus sextupole magnets and octupole magnets respectively.

The ADM cell includes a module for generating micro-bunching in the middle and two matching modules in the head and tail. In Section II, we have mentioned that in order to generate a higher bunching factor, a specific twiss parameter is required at the entrance of the first dipole. In order to cancel the vertical dispersion generated by the dogleg and return the vertical plane to the standard plane, we need to add two dipoles and one dogleg to the structure shown in Fig. 1. In addition, some quadrupoles are needed to match the twiss parameters of this cell with the parameters in the midpoint of the straight section of the ring. The optical functions of the ADM cell are shown in Fig. 7(a), and the complete layout of the horizontal viewing is shown in Fig. 7(b).

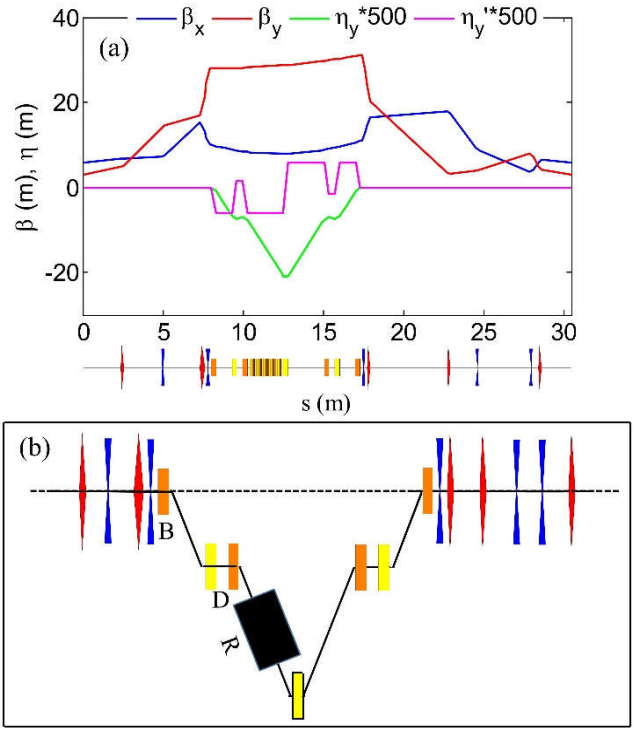

Fig. 7. (a) Optical functions for the ADM cell. (b) The complete layout of the horizontal viewing. The blue, red lens indicate defocus quadrupole and focus quadrupole magnets respectively. Black rectangle

represent radiator. Yellow rectangles indicate normal dipoles, and red rectangles indicate RBs.

The electron beam is deflected downward when passing through the first vertical dipole, then passes through the first dogleg, generates the coherent synchrotron radiation in the slant downward direction through the undulator, and then passes through a vertical dipole with a positive deflection angle twice that of the first dipole, so that the electron beam deflects upward, and then returns to the standard plane through a symmetrical structure. The total length of the ADM cell is about 30 m. Corresponding to this cell is the long drift cell for injection. The 36 arcs of the ring are connected to the ADM cell and long drift cell. The layout of the entire ring is shown in Fig. 8. The horizontal beta function in the midpoint of the long drift for injection is about 18.5 m, this cell consists of six focus quadrupole magnets and four defocus quadrupole magnets. Table II presents the main parameters for our designed lattice. Here, H and V denote the horizontal and vertical directions, respectively. The dipole magnets in ADM cell are deflected in the vertical direction, and the undulator is in a place where the vertical dispersion is relatively large, which contribute to the vertical emittance. The vertical emittance of the whole ring is 20 pm-rad using the beam-envelope matrix method [37].

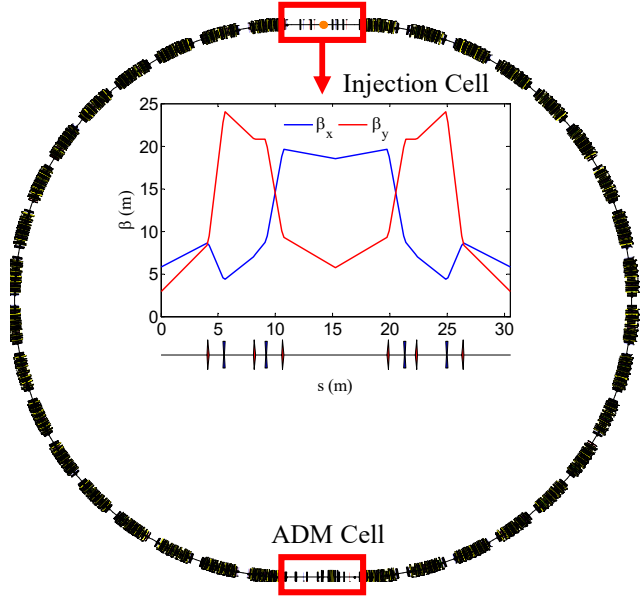

Fig. 8. Layout of the entire ring lattice.

TABLE II. Main Parameters in the designed lattice

| Parameters | Values |
| --- | --- |
| Energy [GeV] | 3.5 |
| Circumference [m] | 900 |
| Emittance (H, V) [pm-rad] | (20, 20) |
| Betatron tune (H, V) | (115.16, 43.27) |
| Energy spread | $1.05 \times 10^{-3}$ |
| Momentum compaction factor | $-1.5 \times 10^{-5}$ |
| Natural chromaticity (H, V) | (-258, -116) |
| Corrected chromaticity (H, V) | (0, 0) |
| Damping time (x, y, z) [ms] | (11.0, 13.7, 7.8) |
| Energy loss per turn [MeV] | 1.5 |
| RMS bunch length [mm] | 1.8 |
| RF voltage [MV] | 2.15 |
| Length of straight section [m] | 4.7 ×34 + 30 ×2 |
| Beta function in straight section (H, V) [m] | (5.8, 2.9) |

Optimization of non-linear optics is the enormous challenge for an ultra-low emittance lattice in order to provide sufficient dynamic acceptance for injection while strong non-linearities are introduced by the sextupole magnets needed for correction of natural chromaticity. It can be seen from Fig. 8 that the entire ring does not have any periodicity, which brings great challenges to nonlinear optimization.

The non-linear design strategy is to realize the MBA arc as a higher order achromat starting from the unit cell with two chromatic sextupole families, the cell tunes are chosen so that all resonances up to second order in sextupole strength cancel over seven unit cells (including two match cells) [36]. Then the tune footprint is tailored to fit between major resonances by using small octupoles, where three families of dispersion-free octupoles in the matching section mainly control the three amplitude dependent tune shifts (ADTS), while two families of dispersive octupoles in the arc mainly control the two second-order chromaticities. The sextupole magnets in unit cell and match cell are in different families. There are a total of 9 optimization parameters here, which are 4 families of sextupole magnets and five families of octupole magnets. Directly tracking dynamic aperture (DA) and local momentum apertures (MA) is a very time-consuming practice, but it is remarkably accurate [38]. We optimize the nonlinearity by optimizing the resonance drive terms (RDT) [39]. Multi-objective particle swarm optimization (MOPSO) [40] has been used for optimization of RDT. Two second-order chromaticity terms and three ADTS are optimized while maintaining the corrected chromaticity to 0. The variation of fractional part tune with energy deviations after MOPSO is depicted in Fig. 9(a). After optimizing chromatic and resonance driving terms, we evaluate the dynamic aperture of designed lattice. Fig. 9(b) shows the results of dynamic aperture tracking in six-dimensional (6D) for different energy deviations in the middle of injection section. The dynamic aperture has been calculated by ELEGANT code, and the particles have been tracked for 2000 turns. The reduction of the off-momentum dynamic aperture is very severe, and the optimization of off-momentum dynamic aperture of this lattice without periodicity is a very difficult problem to solve.

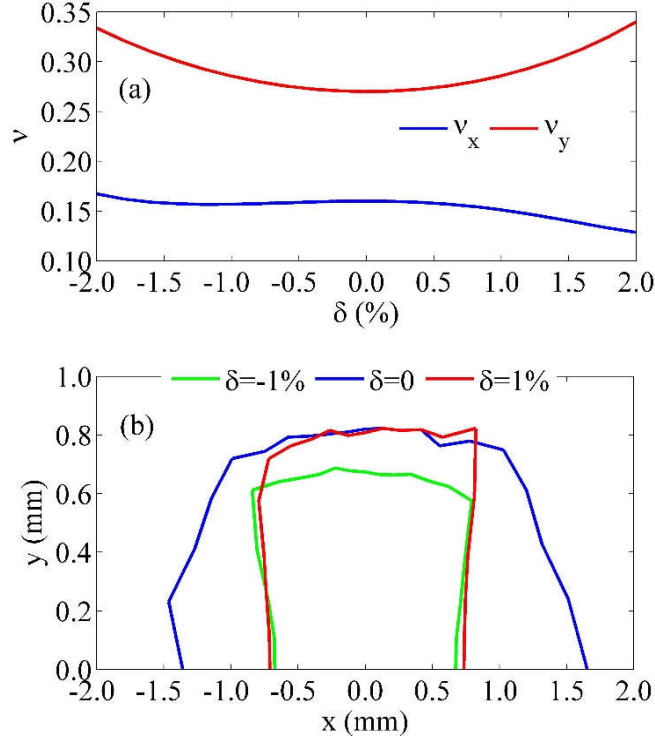

Fig. 9. (a) The variation of fractional part horizontal and vertical tune with energy deviations. (b) The results of DA tracking in 6D for different energy deviation in the middle of injection section.

## IV. Beam distortion and Repetition rate

### A. Nonlinear optimization and beam distortion

The octupole magnets can effectively adjust the second order chromaticities and amplitude dependent tune shifts. Table III shows two second-order chromaticities and three ADTS before and after octupole magnets optimization.

TABLE III. RDT before and after octupole magnets optimization

| Parameters | Before Optimization | After Optimization |
|---|---|---|
| h11002 | -483 | -135 |
| h00112 | 350 | 307 |
| h22000 | $1.77 \times 10^5$ | $4.66 \times 10^5$ |
| h11110 | $2.18 \times 10^6$ | $7.88 \times 10^4$ |
| h00220 | $4.53 \times 10^6$ | $1.84 \times 10^4$ |

From Equation 5 in Ref. [16], we obtain the energy-modulated transport matrix as:

$$M1 = R_D \cdot R_M \cdot R_B = \begin{bmatrix} 1+hb\eta & L & h\eta & \eta - Lb \\ 0 & 1 & 0 & -b \\ b(1+h\xi_D) & \eta & 1+h\xi_D & \xi - \eta b \\ hb & 0 & h & 1 \end{bmatrix}, \quad (1)$$

where $b$ is bending angle of the dipole, $h = k_s \Delta\gamma/\gamma$, $k_s$ is the wave number of the seed laser, $\Delta\gamma$ is the energy modulation amplitude induced by seed laser and $\gamma$ is the relativistic parameter for the mean beam energy. $L = L_M + L_D$, $L_M$ is the length of the modulator and $L_D$ is the length
11

of the dispersion section. $\xi = \xi_M + \xi_D$, $\xi_M$ is the momentum compaction generated in the undulator and $\xi_D$ is the momentum compaction generated in the dogleg. $\eta$ is the dispersion generated in the dogleg. If without energy modulation, the transport matrix is

$$M2 = R_D \cdot R_B = \begin{bmatrix} 1 & L_D & 0 & \eta - L_D b \\ 0 & 1 & 0 & -b \\ b & \eta & 1 & \xi_D - \eta b \\ 0 & 0 & 0 & 1 \end{bmatrix}, \quad (2)$$

Comparing $M1$ and $M2$, it can be find out that R11 of $M1$ get and additional term $hb\eta$ which is related to the energy modulation. It will distort transverse phase space when the electron beam interacts with the laser in the undulator. The beam distributions in the vertical plane after energy modulation is observed in the simulation, which is shown in Fig.10. We used 10 k particles and ELEGANT code in the simulation. Turn 0 represents the initial undisturbed vertical phase space, and blue dot represents the vertical phase space of the first turn after energy modulation. The 600th and 6000th turn are the vertical phase space of natural damping after one modulation. Case (a) and case (b) denote before and after octupole magnets optimization respectively.

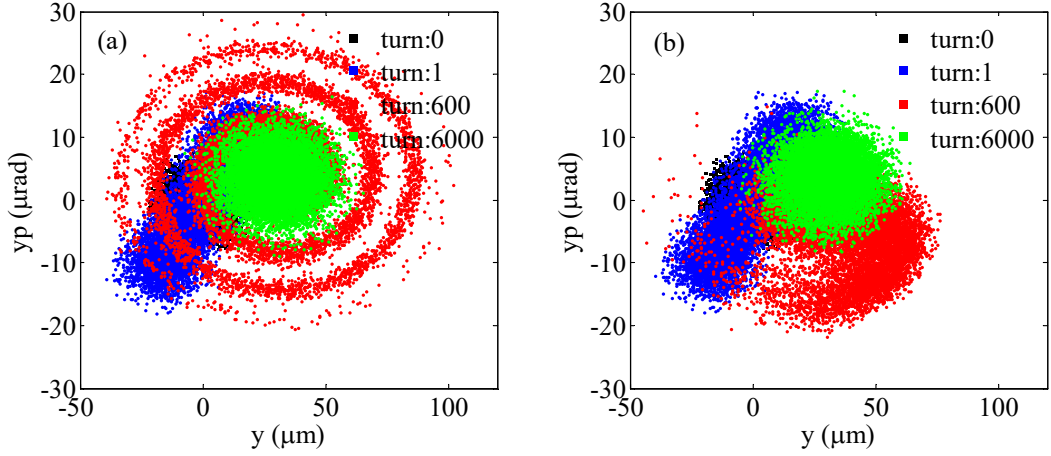

Fig. 10. Beam distributions at the vertical phase space (a) before optimization and (b) after optimization for turn 0 represents the initial undisturbed and turn 1, 600 and 6000 turns represent natural damping after one modulation.

It can be seen from Fig. 10 that the vertical phase space of the 600th turn for case (a) and case (b) are different. The area of the phase space that occupied by the particles represents the emittance. According to the Liouville's theorem the emittance is a constant after beam is modulated. However, the nonlinearity causes the distortion of phase space, which leads to the false imagination of the increasing of the emittance when calculated by the statistical definition $\varepsilon_y = \sqrt{\langle y^2 \rangle \langle y'^2 \rangle - \langle y \cdot y' \rangle^2}$. We call this emittance as the RMS projected emittance [41]. It is necessary to take care of the RMS projected emittance because the light source station receives far-field radiation, the brightness will decrease when the projected emittance blow up. The emittance we analyze below refers to the RMS projected emittance. In Fig.10 (a), the vertical phase space at the 600th turn is like a winding shape, this phase space distortion appears as an increasing of vertical emittance. The phase space distortion comes from accumulation of turn by turn nonlinear effects. Lower the nonlinearity smaller the emittance growth. Fig. 11 shows the vertical emittance with the number of turns after one energy modulation before (case a) and after (case b) octupole magnets optimization.

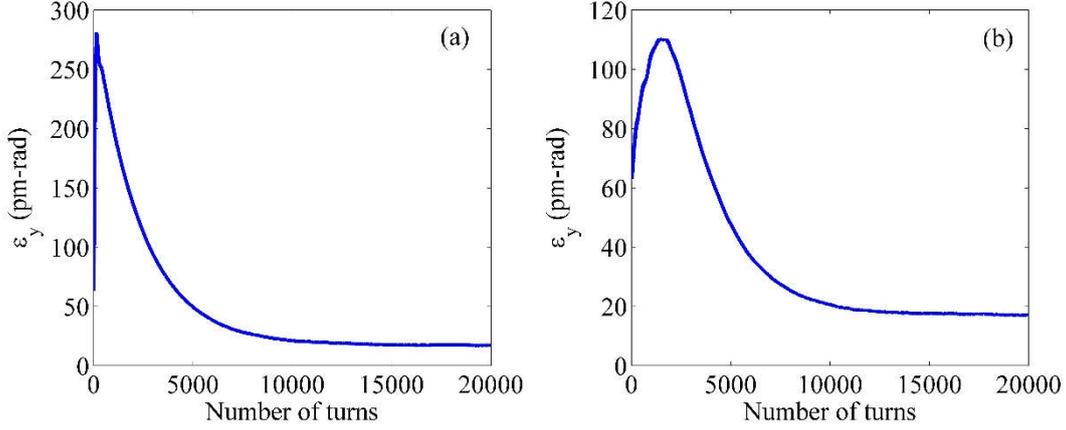

Fig. 11. Variation of vertical emittance with number of turns: case (a) before optimization and case (b) after optimization.

After beam is modulated, it will experience radiation damping process. The next modulation will act after the beam damping to an equilibrium distribution. The increased vertical emittance is mainly caused by the energy modulation. For case (a), the vertical emittance increased very quickly. At the 150th turn, it reached its maximum value, which is about 280 pm-rad. At this time, the vertical phase space was very diffuse. At 6000th turn, the vertical emittance was damped to about 38 pm-rad. The damping time of the first 6000 turns of case (a) is about 8.8 ms ($\tau_{a1}$). Interesting enough, $\tau_{a1}$ is approximately the longitudinal damping time. After then, the damping became slower and slower. At 11000th turn, the vertical emittance was damped to equilibrium, about 20 pm-rad. The damping time between 6000 and 11000 turns is about 23.4 ms ($\tau_{a2}$). For case (b), the vertical emittance increases slower. At 1800th turn, the vertical emittance increases to a maximum value of about 110 pm-rad, and at 6000th turn, damping to about 37 pm-rad. The damping time of the first 6000 turns of case (b) is about 11.6 ms ($\tau_{b1}$), $\tau_{b1}$ is approximately the vertical damping time. At 10000th turn, the vertical emittance damping to equilibrium, approximately 20 pm-rad. The damping time between 6000 and 10000 turns is about 19.5 ms ($\tau_{b2}$). For case (a), the phase space distortion is due to the non-linearity is too strong, and the energy modulation is transferred to the vertical direction. At this point, the damping time of vertical emittance of the first 6000 turns is approximately the longitudinal damping time. For case (b), the damping time of the vertical emittance is determined by the damping time of the vertical direction of about 11.6 ms. From Fig. 11(b), we can see that at 10000 turns, the vertical emittance reaches a balanced state. The entire ring circumference is about 900 m. When using 500MHz RF, the corresponding bucket number is 1500. Among which 300 buckets are filled each with 1.5 nC charge. The total average current is 150 mA. With this filling pattern, we modulate the bunches in turns to produce coherent radiations, the repetition rate can reach 10 kHz.

## B. Brightness and bandwidth

The unique advantage of ADM is producing high intensity radiation pulse with ultra-narrow bandwidth. To illustrate the possible performance of ADM, 3D simulations for the 6[th] harmonic (44.3 nm) radiation with GENESIS [42] have been performed with the bunched electron beam from ELEGANT. The radiator is a 3 m long helical undulator with period of 8 cm. Fig. 12 shows simulation results for output radiation pulse and single-shot spectrum by the end of radiator. One can find the peak power is over 1MW (~$10^{12}$ phs/pulse). The spectral bandwidth (FWHM) is only 0.3 meV, which is quite close to the Fourier transform limit. Considering a pulse repetition rate of 10 kHz, the spectral brightness is calculated to be about $3\times10^{24}$ phs/s/mm$^2$/mrad$^2$/0.1%BW, over two orders of magnitude higher than spontaneous emission with the same electron beam and undulator in the general fourth generation light source. Also, for the flux, it reaches $10^{16}$ phs/s. As the radiation is extremely narrow banded, monochromator is not necessary in the beam line which will allow full usage of the flux.

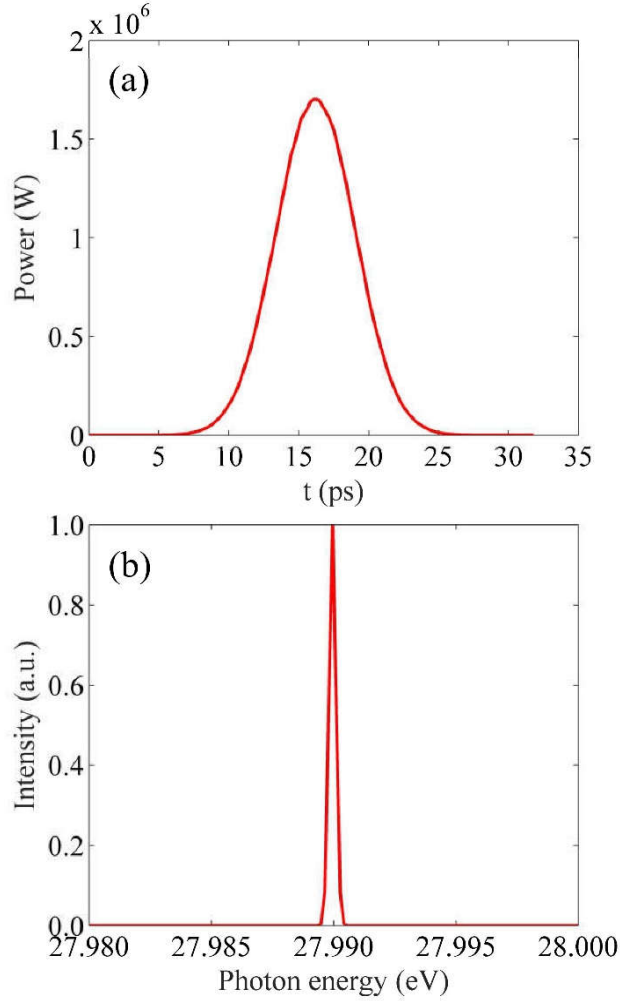

Fig. 12. (a) Power of the radiation pulse. (b) Spectral of the radiation pulse

## V. DISSCUSSIONS AND CONCLUSIONS

In this work, the novel ADM technique is used to enhance the coherent micro-bunching. Numerical simulations demonstrated that this scheme can increase the brightness 2 orders higher than the spontaneous radiation in the storage ring. The 6$^{th}$ harmonic (44.3 nm) of 260 nm seeding laser is taken as an example for brightness calculation. At this photon energy there're some important applications such as super-high energy resolution Angle-Resolved Photoemission Spectroscopy (ARPES) [43]. Without remodulation, the repetition rate can reach 10 kHz which is strongly depends on the damping time and the storage ring nonlinear term optimization.

Owing to the long bunch length and the flat energy chirp of the electron beam, the radiation will be extremely narrow banded beyond the capability of a monochromator that could provide. As beam in the storage ring is highly stable, the radiation will also be very stable, which is preferred by the users in any case.

# ACKNOWLEDGEMENTS

The authors are deeply grateful to Alex Chao (Tsinghua University) and Weishi Wan (ShanghaiTech University) for guidance of the study. Thanks also to Zhenghe Bai, Tong Zhang and Penghui Yang at the University of Science and Technology of China (USTC) for help discussions on the storage ring lattice design. This work is supported by the National Natural Science Foundation of China (No. 11975300).